\documentclass[11pt,british,abstracton]{scrartcl}
\usepackage{ae}
\usepackage{aecompl}
\usepackage[]{fontenc}
\usepackage[latin1]{inputenc}
\usepackage{geometry}
\usepackage{subfigure}
\usepackage{amsmath}
\usepackage{graphicx}
\usepackage{amssymb}

\makeatletter

\providecommand{\tabularnewline}{\\}

\usepackage{babel}
\makeatother
\begin{document}

\titlehead{\hfill{}TUM-HEP-611/05}

\title{Scaling Laws for the Cosmological Constant}

\author{Florian Bauer%
\thanks{Email: \texttt{fbauer@ph.tum.de}%
}\\
 \\
\emph{\large Physik Department T30d, Technische Universität München,}\\
\emph{\large James-Franck-Straße, 85748 Garching, Germany}}

\date{{\large 1 December 2005}}

\maketitle
\begin{abstract}
We study the expansion of the universe at late times in the case that
the cosmological constant obeys certain scaling laws motivated by
renormalisation group running in quantum theories. The renormalisation
scale is identified with the Hubble scale and the inverse radii of
the event and particle horizon, respectively. We find de~Sitter solutions,
power-law expansion and super-exponential expansion in addition to
future singularities of the Big Rip and Big Crunch type. 
\end{abstract}
\renewcommand{\textrm}{\text}

\newcommand{\kl}[1]{{\textstyle #1}}

\renewcommand*{\caplabelfont}{\sffamily\bfseries\small} \renewcommand*{\capfont}{\normalfont\small}

\section{Introduction}

The recently observed~\cite{Riess:1998cb,Perlmutter:1998np,Spergel:2003cb,Tegmark:2003ud,Boughn:2004zm,Hannestad:2005fg}
accelerated expansion of the universe has opened new ways to think
about the connection between cosmology and quantum physics since an
entire classical origin for the acceleration seems improbable. Quantum
theories, on the other hand, potentially yield a lot of sources for
dark energy (DE), the energy form that is responsible for the cosmic
acceleration \cite{Peebles:2002gy,Sahni:2004ai,paddy}. One of the
most prominent candidates for DE is a positive cosmological constant
(CC) because it appears already in Einstein's equations of classical
general relativity. In this context, it can be treated as a perfect
fluid with a constant energy density~$\Lambda>0$ and an equation
of state (EOS) $\omega=p/\Lambda=-1$ that corresponds to a negative
pressure~$p=-\Lambda$. Furthermore, on the quantum level the CC
can be interpreted as vacuum energy, however, its absolute value has
not been predicted reliably in any theory yet, and naive estimates
are far too large~\cite{Weinberg:1988cp}. 

In this work we investigate the late-time cosmological evolution in
a special class of cosmologies that are based on general relativity
with a time-dependent CC and Newton's constant. Classically both constants
are truly time-independent, but quantum effects may induce a scale-dependence
in the form of renormalisation group equations~(RGE) for these constants.
For instance, quantum field theory~(QFT) on curved space-times~\cite{QFTonCurve,RGE-LamG}
and quantum gravity~\cite{QEG,time-LamG-QG-1,time-LamG-QG-2} provide
a framework for the renormalisation group running. However, the corresponding
RGEs are not unique and the identification of the renormalisation
scale~$\mu$, on which they depend, is not fixed by the theories.
For phenomenological reasons, we should request that the scale~$\mu$
does not change too much over cosmological time scales. In contrast
to DE scenarios with a time-dependent EOS~\cite{quintessence,time-dep-EOS},
here, the EOS of the CC is still exactly~$\omega=-1$. However, one
obtains an effective time-dependent EOS~\cite{Eff-EOS} due to the
non-standard scaling of the matter energy density with the cosmological
time.

The paper is organised as follows: in Sec.~\ref{sec:Scaling-laws}
we describe three scaling laws for the CC. Two of them are based on
QFT on curved space-times, whereas the third one appears in the context
of quantum Einstein gravity (QEG) at late cosmological times. In the
latter case we also consider the running of Newton's constant. The
identification of the renormalisation scale~$\mu$ with some cosmologically
relevant quantity is done in Sec.~\ref{sec:Choice-scale}. We consider
three possible choices for~$\mu$: the Hubble scale~$H$, the (inverse)
radius~$R$ of the cosmological event horizon and finally the (inverse)
radius~$T$ of the particle horizon. In Sec.~\ref{sec:Time-Einstein}
we discuss how to solve Einstein's equations on a Robertson-Walker
background when the CC and Newton's constant depend on the cosmological
time. The main part of this work can be found in Sec.~\ref{sec:Late-time},
where we investigate the late-time evolution of the universe for all
combinations of scaling laws and scale choices. Since it is not always
possible to obtain exact or approximate solutions, some of the given
results are numerical. Finally, we present our conclusions in Sec.~\ref{sec:Conclusions}.

\section{\label{sec:Scaling-laws}Scaling laws}

According to QFT on curved space-times~\cite{QFTonCurve,RGE-LamG}
the CC and Newton's constant are subject to renormalisation group
running like the running of the fine-structure constant in quantum
electrodynamics. The RGE for the CC, corresponding to the vacuum energy
density~$\Lambda$, depends crucially on the considered quantum fields
and their masses. For this kind of running we neglect the scale-dependence
of Newton's constant in this work.

In the following we discuss two different RGEs, the first one follows
directly from the 1-loop effective action and is given by $\mu\frac{\textrm{d}\Lambda}{\textrm{d}\mu}=A_{0}$
in the $\overline{\textrm{MS}}$-scheme, where quantum fields with
masses~$M$ determine the coefficient~$A_{0}\sim\mp M^{4}$. Note
that the sign of~$A_{0}$ depends on the spins of the fields, see,
e.g., Ref.~\cite{Bauer:2005rp} for the exact expression. Upon integration
one obtains in the case of constant masses\begin{equation}
\frac{\Lambda(\mu)}{\Lambda_{0}}=1-q_{1}\ln\frac{\mu}{\mu_{0}},\,\,\,\, q_{1}\sim\mp\frac{M^{4}}{\Lambda_{0}},\label{eq:RGE1}\end{equation}
where~$\Lambda_{0}$ denotes the vacuum energy density today, when
the renormalisation scale~$\mu$ has the value~$\mu_{0}$. The scaling
law~(\ref{eq:RGE1}) has a big disadvantage since it has been derived
in a renormalisation scheme that is usually associated with the high-energy
regime~\cite{decRGE}. It is therefore not known to what extend it
can be applied at late times in cosmology. In addition, to avoid conflicts
with observations, the field content has to be fine-tuned to obtain
$|q_{1}|\le O(1)$, which we assume for the rest of this work. 

The second scaling law follows from a RGE that shows a decoupling
behaviour. Considering only the most dominant terms at low energy,
the corresponding $\beta$-function for~$\Lambda$ is given by $\mu\frac{\textrm{d}\Lambda}{\textrm{d}\mu}=A_{1}\mu^{2}$,
where $A_{1}\sim\pm M^{2}$ is set by the masses~$M$ and the spins
of the fields. RGEs of this kind have been studied extensively in
the literature \cite{RGE-LamG}. Assuming constant masses and $\mu_{0}$
to be of the order of today's Hubble scale~$H_{0}$ one obtains\begin{equation}
\frac{\Lambda(\mu)}{\Lambda_{0}}=L_{0}+L_{1}\frac{\mu^{2}}{\mu_{0}^{2}},\,\,\,\, L_{1}\sim\pm\frac{M^{2}}{M_{\textrm{P}}^{2}},\label{eq:RGE2}\end{equation}
where~$L_{0}:=1-L_{1}$ and~$M_{\textrm{P}}$ denotes the Planck
mass today. Here, the running of~$\Lambda$ is suppressed since~$|L_{1}|\ll1$
for sub-Planckian masses~$M$.

The last scaling laws come from the RGEs in QEG \cite{QEG,time-LamG-QG-1}.
In this framework the effective gravitational action becomes dependent
on a renormalisation scale~$\mu$, which leads to RGEs for~$\Lambda$
and~$G$. An interesting feature is the occurrence of an ultraviolet~(UV)
fixed-point~\cite{QEG-Fixpoint} in the renormalisation group flow
of the dimensionless quantities%
\footnote{In this work~$\Lambda$ denotes the vacuum energy density corresponding
to a CC, whereas in many articles about QEG~$\Lambda$ is the CC
itself.%
}~$\Lambda\mu^{-4}$ and~$G\mu^{2}$ at very early times in cosmology
corresponding to~$\mu\rightarrow\infty$. Motivated by strong infrared
(IR) effects in quantum gravity one has proposed that there might
also exist an infrared fixed-point in QEG~\cite{QEG-IR} leading
to significant changes in cosmology at late times%
\footnote{However, it was argued in Ref.~\cite{QEG-NoIR} that the region of
strong IR effects can never be reached.%
}, where $\mu\rightarrow0$. If this were true, one would obtain the
RGEs\begin{equation}
\frac{\Lambda}{\Lambda_{0}}=\frac{\mu^{4}}{\mu_{0}^{4}},\,\,\,\,\frac{G}{G_{0}}=\frac{\mu_{0}^{2}}{\mu^{2}},\label{eq:RGE3}\end{equation}
 where $\mu_{0}$ corresponds to $\Lambda=\Lambda_{0}$ and~$G=G_{0}$.
In the epoch between the UV and IR fixed-points,~$\Lambda$ and~$G$
vary very slowly with~$\mu$, and we will treat them as constants
in this region. Therefore, we assume that the scaling laws~(\ref{eq:RGE3})
are valid from today until the end of the universe.

\section{\label{sec:Choice-scale}Choice of the scale}

In order to study the effects on the cosmological expansion due to
the scaling laws of Sec.~\ref{sec:Scaling-laws}, we have to define
the physical meaning of the renormalisation scale~$\mu$. Unfortunately,
the theories underlying the scaling laws do not determine the scale
explicitly%
\footnote{In the framework of Refs.~\cite{G-von-H}, Newton's constant was
found to depend explicitly on the Hubble scale.%
}, apart from the usual interpretations as an (IR) cutoff or a scale
characterising the physical environment (temperature, external momenta).
In our cosmological setting we will investigate three different choices
for~$\mu$, given by the Hubble scale~$H$, the inverse radius~$R^{-1}$
of the cosmological event horizon, and finally the inverse radius~$T^{-1}$
of the particle horizon.

\begin{figure}
\begin{centering}\includegraphics[clip,width=0.48\columnwidth,keepaspectratio]{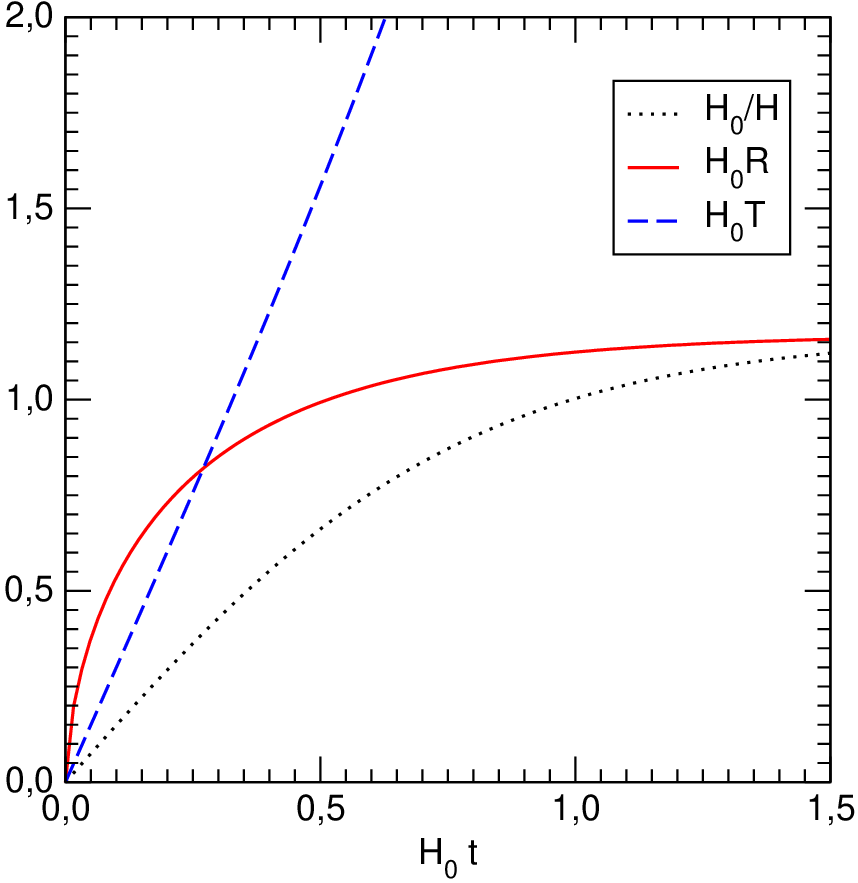}\hfill{}\includegraphics[clip,width=0.48\columnwidth,keepaspectratio]{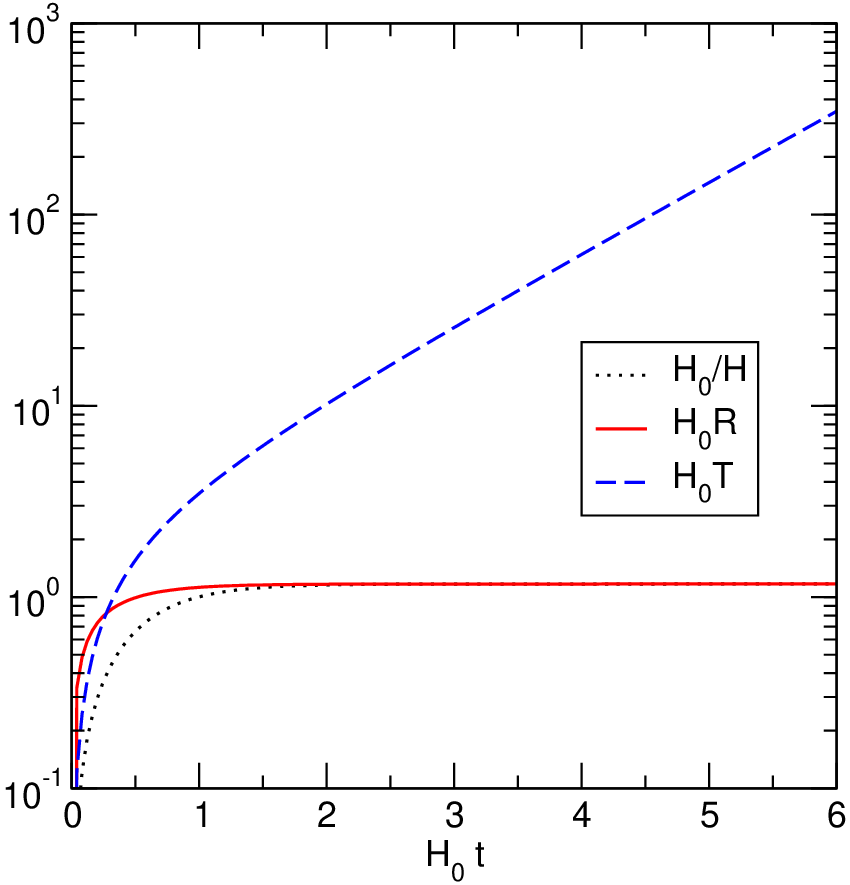}\par\end{centering}

\caption{\label{cap:ScalesLCDM}In a spatially flat $\Lambda$CDM universe
with constant~$\Lambda$ and~$G$ the radii of the event horizon~$R$
and the Hubble horizon~$H^{-1}$ approach the same constant value
for~$t\rightarrow\infty$. At early times the particle horizon radius~$T$
grows linearly with time and exponentially with time for~$t\rightarrow\infty$.
At the time~$t_{0}=0,99H_{0}^{-1}$ (today) the relative vacuum energy
is given by~$\Omega_{\Lambda0}=0,73$. }
\end{figure}

For the purposes of this work we consider a spatially flat Robertson-Walker
metric given by the line element\begin{equation}
\textrm{d}s^{2}=\textrm{d}t^{2}-a^{2}(t)\textrm{d}\vec{x}^{2},\label{eq:RWmetric}\end{equation}
where the scale factor~$a(t)$ depends only on the cosmological time~$t$
and not on the spatial coordinates~$\vec{x}$. While the Hubble scale\begin{equation}
H(t):=\frac{\dot{a}(t)}{a(t)}\label{eq:Scale1}\end{equation}
 describes the actual expansion rate of the universe, the horizon
scales~$R$ and~$T$ describe the cosmological evolution of the
future and the past, respectively. The radius~$R$ of the cosmological
event horizon corresponds to the proper distance that a (light) signal
can travel when it is emitted by a comoving observer at the time~$t$:
\begin{equation}
R(t):=a(t)\int_{t}^{\infty}\frac{\textrm{d}t^{\prime}}{a(t^{\prime})}.\label{eq:Scale2}\end{equation}
In the case that the universe comes to an end within finite time the
upper limit of the integral has to be replaced by this time. Similar
to the event horizon of a black hole, the cosmological event horizon
exhibits thermodynamical properties like the emission of radiation
with the Gibbons-Hawking temperature~\cite{horizon-thermo}. The
analogue to~$R$ is the particle horizon radius~$T$, which is given
by the proper distance that a signal has travelled since the beginning
of the world ($t=0$):\begin{equation}
T(t):=a(t)\int_{0}^{t}\frac{\textrm{d}t^{\prime}}{a(t^{\prime})}.\label{eq:Scale3}\end{equation}
In a simple cosmological model, where the universe begins at~$t=0$
and then evolves according to the scale factor~$a(t)\propto t^{n}$
with~$0<n<1$, the particle horizon radius reads $T(t)=t/(1-n)$.
Now we assume that at some time~$t_{0}$ a de~Sitter phase sets
in, which corresponds to a scale factor~$a(t)\propto\exp(H_{0}(t-t_{0}))$
for~$t>t_{0}$. This leads to a radius function~$T$ that grows
exponentially with~$t$ at late times:\begin{equation}
T(t>t_{0})=\left(T(t_{0})+\frac{1}{H_{0}}\right)\exp(H_{0}(t-t_{0}))-\frac{1}{H_{0}}.\label{eq:T-dS}\end{equation}
Since the asymptotic behaviour of the scales~$H$,~$R$ and~$T$
plays a major role in this work, it is plotted in Fig.~\ref{cap:ScalesLCDM}
for a universe with dust-like matter and~$\Lambda$ and~$G$ being
positive and constant.%
 Note that if one of the horizon radii diverges, one says that the
horizon does not exist. The geometric meaning of the scales given
here has been discussed, e.g., in Ref.~\cite{horizons}, and some
applications in the context of DE can be found in Refs.~\cite{horizon-DE}.

\section{\label{sec:Time-Einstein}Time-dependent~$\boldsymbol{\Lambda}$
and~$\boldsymbol{G}$}

Applying the scaling laws of Sec.~\ref{sec:Scaling-laws} together
with the identification of the scale~$\mu$ with one of the cosmological
scales of Sec.~\ref{sec:Choice-scale}, we obtain time-dependent
functions~$\Lambda(t)$ and~$G(t)$. Einstein's equations $G^{\alpha\beta}=8\pi G(\Lambda g^{\alpha\beta}+T^{\alpha\beta})$
and the Einstein tensor~$G^{\alpha\beta}$ remain unmodified apart
from the additional time-dependence in~$\Lambda$ and~$G$. For
the late-time epoch we consider the spatially flat metric~$g^{\alpha\beta}$
following from Eq.~(\ref{eq:RWmetric}), and the energy-momentum
tensor~$T^{\alpha\beta}$ describes a perfect fluid with energy density~$\rho$
and pressure~$p$. Additionally, we introduce the variable~$Q:=(1+3\omega)/2$
which is fixed by the EOS~$\omega=p/\rho$ of the fluid. Finally,
the evolution of the scale factor~$a(t)$ is given by Friedmann's
equations\begin{equation}
\frac{\dot{a}^{2}}{a^{2}}=\frac{8\pi}{3}G(\Lambda+\rho),\,\,\,\,\frac{\ddot{a}}{a}=\frac{8\pi}{3}G(\Lambda-Q\rho),\label{eq:Friedmann}\end{equation}
which have to agree with the Bianchi identities\[
0=G_{\,\,\,\,\,\,;\beta}^{\alpha\beta}=[G\Lambda g^{\alpha\beta}+GT^{\alpha\beta}]_{;\beta},\]
that imply\[
\dot{G}(\Lambda+\rho)+G(\dot{\Lambda}+\dot{\rho}+2H(Q+1)\rho)=0.\]
For constant~$\Lambda$ and~$G$ the last equation can be integrated
to yield the usual scaling law for the matter energy density~$\rho\propto a^{-2(Q+1)}$.
In general, this is not possible when~$\Lambda$ and~$G$ depend
on time due to the energy transfer between~$\Lambda$ and~$\rho$.
It therefore implies an effective interaction between the gravitational
sector and matter, which is not part of the original Lagrangian. In
this sense it should be compared with gravitational particle production~\cite{Parker:1968mv}
resulting also from the interplay of gravity with quantum physics.

To determine the scale factor~$a(t)$ we combine both Friedmann's
equations (\ref{eq:Friedmann}) resulting in\begin{equation}
K_{0}F(t)=\frac{G(t)}{G_{0}}\frac{\Lambda(t)}{\Lambda_{0}}\,\,\,\,\textrm{with}\,\,\,\, F(t):=\frac{\ddot{a}}{a}+Q\left(\frac{\dot{a}}{a}\right)^{\!2}=\dot{H}+(Q+1)H^{2},\label{eq:MainEqu}\end{equation}
where the constant~$K_{0}:=[\Omega_{\Lambda0}(Q+1)H_{0}^{2}]^{-1}$
is fixed by today's values of the Hubble scale~$H_{0}$, the relative
vacuum energy density~$\Omega_{\Lambda0}$ and the matter EOS~$Q$.

\section{\label{sec:Late-time}Late-time evolution}

In this section we study the late-time evolution of the universe with
variable~$\Lambda$ and~$G$, thereby assuming from today on the
validity of the scaling laws (\ref{eq:RGE1})--(\ref{eq:RGE3}) and
the correct identification of the renormalisation scale~$\mu$ with
the scales (\ref{eq:Scale1})--(\ref{eq:Scale3}). Our aim is to determine
in all nine cases the possible final states of the universe. This
depends, of course, on the choice of parameters, but we restrict ourselves
to parameter values which comply vaguely to current observations.
Today, at the cosmological time~$t_{0}=0,99H_{0}^{-1}$ we fix the
initial values~$H_{0}$ and~$\Omega_{\Lambda0}=0,73$ by observations.
Furthermore, the initial value~$T_{0}$ of the particle horizon radius
would be fixed if the past cosmological evolution was known from the
Big Bang on. In contrast to this, the value~$R_{0}$ of today's event
horizon radius depends on the future cosmological evolution and is
treated here as a free parameter. In the following we derive some
properties of the solutions analytically, in particular, we study
the stability of (asymptotic) de~Sitter solutions and the occurrence
of future singularities~\cite{Future-Sing}, where the scale factor
or one of its derivatives diverge within finite time. For simplicity
we denote a Big Rip or a Big Crunch by the lowest order divergent
derivative that is positive or negative, respectively. Since some
combinations of scaling laws and renormalisation scales lead to complicated
equations we derive in some cases only approximate or numerical statements. 

In order to solve Eq.~(\ref{eq:MainEqu}) numerically, we have to
remove the integrals in the definitions (\ref{eq:Scale2}), (\ref{eq:Scale3})
of~$R$ and~$T$ by a differentiation with respect to~$t$. Using
the relations \[
\dot{R}=RH-1,\,\,\,\,\dot{T}=TH+1\]
 an ordinary differential equation for the scale factor can be obtained
and numerically integrated. Afterwards we have to check whether the
functions~$R$ and~$T$, calculated from the numerical solution
of~$a(t)$, agree with~$R$ and~$T$ that follow directly from
Eq.~(\ref{eq:MainEqu}). If they do not match, the numerical solution
has to be discarded. Furthermore, solutions involving a negative matter
energy density are questionable on physical grounds. This happens
when the vacuum energy density~$\Lambda$ becomes greater than the
critical energy density~$\rho_{\textrm{c}}=3H^{2}/(8\pi G)$, which
follows from the first Friedmann equation~(\ref{eq:Friedmann}).

\subsection{\label{sub:A1}$\boldsymbol{\Lambda=\Lambda_{0}(1-q_{1}\ln\frac{\mu}{\mu_{0}})}$,
$\boldsymbol{\mu=H}$}

\begin{figure}
\begin{centering}\includegraphics[clip,width=0.4\columnwidth,keepaspectratio]{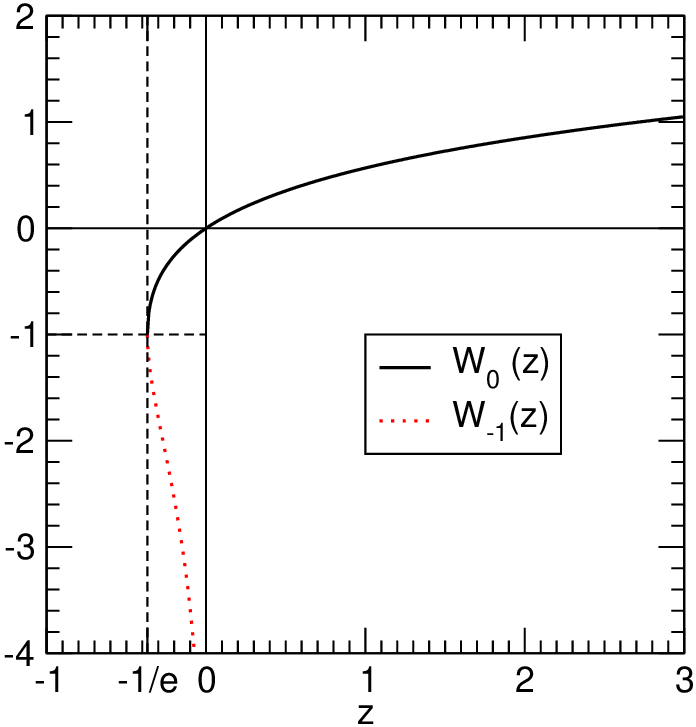}\par\end{centering}

\caption{\label{cap:LambertW}Lambert's $W$-function is the inverse function
of~$z=xe^{x}$. For~$-1/e\le z<0$ it has two real-valued brunches~$W_{0}(z)$
and~$W_{-1}(z)$, and for~$z\ge0$ only~$W_{0}(z)$ is real.}
\end{figure}

\begin{figure}
\begin{centering}\includegraphics[clip,width=1\columnwidth,keepaspectratio]{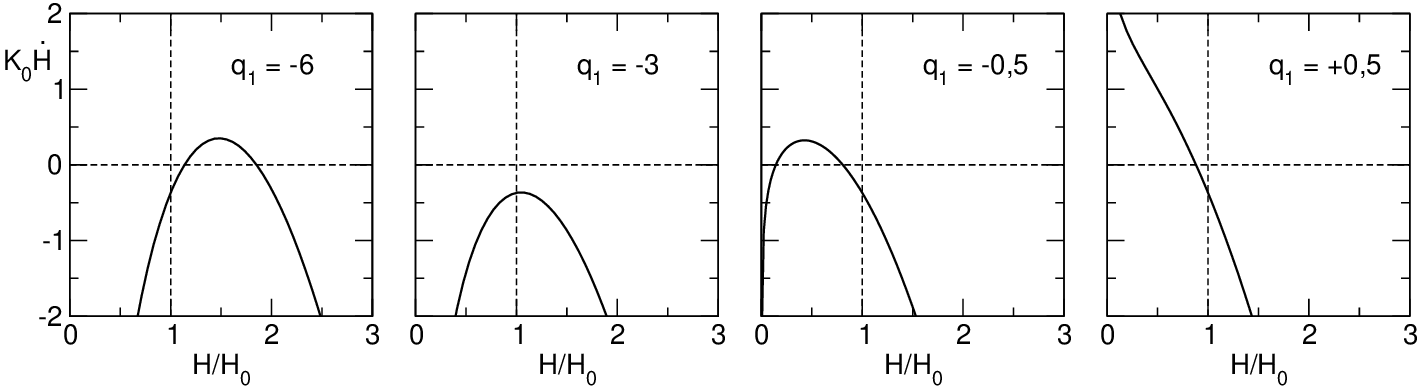}\par\end{centering}

\caption{\label{cap:A1-HdotH}The phase space relation~(\ref{eq:A1-Hdot})
is shown for~$q_{1}=-6;-3$, where the universe ends in a Big Crunch
singularity, and respectively for~$q_{1}=-\frac{1}{2};+\frac{1}{2}$,
which leads to a de~Sitter final state ($\dot{H}=0$). In all cases
the evolution begins at~$H=H_{0}$, where~$\Omega_{\Lambda0}=0,73$
and~$\dot{H}_{0}<0$.}
\end{figure}

For the given scaling law and scale choice Eq.~(\ref{eq:MainEqu})
reads\begin{equation}
K_{0}(\dot{H}+(Q+1)H^{2})=1-q_{1}\ln\frac{H}{H_{0}}.\label{eq:A1-Hdot}\end{equation}
 We first look for asymptotic de~Sitter solutions by applying $\dot{H}=0$
and~$H\rightarrow H_{\textrm{e}}$. Thus the final Hubble scale~$H_{\textrm{e}}$
is given by \begin{equation}
\frac{H_{\textrm{e}}^{2}}{H_{0}^{2}}=\frac{q_{1}}{2}\Omega_{\Lambda0}W_{u}\left(\frac{2}{q_{1}\Omega_{\Lambda0}}e^{2/q_{1}}\right)\!\!.\label{eq:A1-He}\end{equation}
Here, $W_{u}(z)$ with~$u=0,-1$ denotes one of the two real-valued
branches of Lambert's \mbox{W-function}, which is the solution of~$z=xe^{x}$,
see Fig.~\ref{cap:LambertW}.%
 For~$q_{1}>0$ there is always one solution for~$H_{\textrm{e}}$,
and for~$q_{1}<0$ two solutions exist if the argument of~$W_{u}$
in Eq.~(\ref{eq:A1-He}) is greater than~$-e^{-1}$. This means
either~$q_{1}<n_{1}$ or~$q_{1}>n_{2}$ with~$n_{1}:=2/W_{0}(-\Omega_{\Lambda0}e^{-1})$
and~$n_{2}:=2/W_{-1}(-\Omega_{\Lambda0}e^{-1})>n_{1}$. These solution
are stable if\[
\left.\frac{\textrm{d}(K_{0}\dot{H})}{\textrm{d}H}\right|_{H\rightarrow H_{\textrm{e}}}=-\frac{q_{1}}{H_{\textrm{e}}}-2H_{\textrm{e}}K_{0}(Q+1)<0,\]
where we used Eq.~(\ref{eq:A1-Hdot}). For positive~$q_{1}$ this
condition is always fulfilled, whereas for negative~$q_{1}$ it means~$W_{u}(\frac{2}{q_{1}\Omega_{\Lambda0}}e^{2/q_{1}})<-1$,
which follows from Eq.~(\ref{eq:A1-He}). Again, the argument of~$W_{u}$
is constrained yielding~$q_{1}>n_{2}$. Therefore only $q_{1}>n_{2}$,
which includes~$q_{1}>0$, leads to stable de~Sitter solutions.
Using the phase space relation~(\ref{eq:A1-Hdot}), which is plotted
in Fig.~\ref{cap:A1-HdotH}, we conclude that for other values of~$q_{1}$
the cosmological evolution will always end within finite time in a
Big Crunch singularity, where~$H\rightarrow0$ and~$\dot{H}\rightarrow-\infty$.%

\subsection{\label{sub:B1}$\boldsymbol{\Lambda=\Lambda_{0}(L_{0}+L_{1}\frac{\mu^{2}}{\mu_{0}^{2}})}$,
$\boldsymbol{\mu=H}$}

Using this choice for~$\Lambda$ and~$\mu$ Eq.~(\ref{eq:MainEqu})
becomes\[
\frac{\ddot{a}}{a}+H^{2}(Q-L_{1}(Q+1)\Omega_{\Lambda0})=H_{0}^{2}L_{0}\Omega_{\Lambda0}(Q+1),\]
which has the exact solution\[
\frac{a(t)}{a_{0}}=\left[\sinh([Q+1][\Omega_{\Lambda0}L_{0}(1-L_{1}\Omega_{\Lambda0})]^{\frac{1}{2}}H_{0}t)\right]^{n}\]
with\[
n:=[(Q+1)(1-L_{1}\Omega_{\Lambda0})]^{-1}.\]
At early times,~$H_{0}t\ll1$, it describes a power-law expansion
$a(t)\propto t^{n}$, whereas at late times it approaches the de~Sitter
expansion law $a(t)\propto\exp(H_{\textrm{e}}t)$ with the final Hubble
scale~$H_{\textrm{e}}$ given by\[
H_{\textrm{e}}=H_{0}\sqrt{\frac{\Omega_{\Lambda0}L_{0}}{1-L_{1}\Omega_{\Lambda0}}}.\]
Further aspects of this case have been studied, e.g., in Refs.~\cite{RGE-LamG}.

\subsection{\label{sub:C1}$\boldsymbol{\Lambda G=\Lambda_{0}G_{0}\frac{\mu^{2}}{\mu_{0}^{2}}}$,
$\boldsymbol{\mu=H}$}

Following Sec.~\ref{sec:Scaling-laws} we assume that this scaling
law is valid from today on. Here, Eq.~(\ref{eq:MainEqu}) is given
by~$\frac{\ddot{a}}{a}+BH^{2}=0$ with~$B:=Q-(Q+1)\Omega_{\Lambda0}$.
It has an exact power-law solution\[
\frac{a(t)}{a_{0}}=[(1+B)(t-t_{1})]^{1/(B+1)},\,\,\,\, t_{1}=\textrm{const.},\]
that exhibits a constant acceleration~$\ddot{a}a/\dot{a}^{2}=-B>0$
as well as~$\Omega_{\Lambda}=\textrm{const}$. A similar solution
was found in Refs.~\cite{QEG-IR}.

\subsection{\label{sub:A2}$\boldsymbol{\Lambda=\Lambda_{0}(1-q_{1}\ln\frac{\mu}{\mu_{0}})}$,
$\boldsymbol{\mu=R^{-1}}$}

Since this case has been discussed explicitly in Ref.~\cite{Bauer:2005rp},
we skip some details in the following discussion. At first we look
for de~Sitter solutions, where the inverse Hubble scale and the event
horizon radius approach~$R\rightarrow R_{\textrm{e}}=H_{\textrm{e}}^{-1}$
in addition to~$\dot{H}=0$. Then~$K_{0}F=1+q_{1}\ln\frac{R}{R_{0}}$
can be solved for~$R_{\textrm{e}}$ yielding\begin{equation}
R_{\textrm{e}}^{2}=\frac{2}{H_{0}^{2}q_{1}\Omega_{\Lambda0}W_{u}(\frac{2}{R_{0}^{2}H_{0}^{2}q_{1}\Omega_{\Lambda0}}e^{2/q_{1}})}.\label{eq:A2-Re}\end{equation}
For positive~$q_{1}$ there is always a solution for~$R_{\textrm{e}}$,
but it is unstable as can be seen from the following consideration.
We first insert~$R=R_{0}\exp([K_{0}F-1]/q_{1})$ and~$H=\sqrt{K_{0}F/[K_{0}(Q+1)]}$,
which is valid in the de~Sitter limit, into~$K_{0}\dot{F}=q_{1}(H-R^{-1})$.
Then we compute\begin{equation}
\left.\frac{\textrm{d}(K_{0}\dot{F})}{\textrm{d}(K_{0}F)}\right|_{\textrm{dS}}=\frac{q_{1}}{2}[K_{0}^{2}F(Q+1)]^{-\frac{1}{2}}+\frac{1}{R_{0}}\exp\left[\frac{1-K_{0}F}{q_{1}}\right]\!\!,\label{eq:A2-dFdotdF}\end{equation}
which is positive for~$q_{1}>0$ signalling instability. Thus a Big
Crunch singularity ($F,\dot{H}\rightarrow-\infty$, $R\rightarrow0$)
might happen for certain values of initial conditions and parameters,
or there exists no solution. A Big Rip singularity ($F,\dot{H}\rightarrow\infty$,
$R\rightarrow0$) does not exist because it is not compatible with~$R\rightarrow\infty$.

For negative~$q_{1}$ there are two solutions for~$R_{\textrm{e}}$
if the argument of~$W_{u}$ in Eq.~(\ref{eq:A2-Re}) is greater
than~$-e^{-1}$ corresponding to the condition~\[
R_{0}>\sqrt{\frac{2\exp(2/q_{1}+1)}{-q_{1}H_{0}^{2}\Omega_{\Lambda0}}}.\]
 According to Eq.~(\ref{eq:A2-dFdotdF}) it turns out that~$R_{\textrm{e}1}:=R_{\textrm{e}}|_{u=0}$
represents a stable solution and~$R_{\textrm{e}2}:=R_{\textrm{e}}|_{u=-1}$
an unstable one, respectively. A closer inspection of the solutions~$R_{\textrm{e}1/2}$
shows that~$R_{\textrm{e}1}>R_{0}>R_{\textrm{e}2}$ if~$R_{0}^{2}H_{0}^{2}\Omega_{\Lambda0}>1$.
In the case~$R_{0}^{2}H_{0}^{2}\Omega_{\Lambda0}<1$ the parameter
range\[
q_{1}<\frac{2}{W_{0}(-1/(e^{1}q_{3}))}\]
 results to~$R_{0}>R_{\textrm{e}1}>R_{\textrm{e}2}$, whereas\[
q_{1}>\frac{2}{W_{0}(-1/(e^{1}q_{3}))}\]
 leads to~$R_{\textrm{e}1}>R_{\textrm{e}2}>R_{0}$, respectively.
Moreover, all stable de~Sitter solutions are realised except the
last one corresponding to the parameter region ($R_{0}^{2}H_{0}^{2}\Omega_{\Lambda0}<1$,
$q_{1}>2/W_{0}[-1/(e^{1}q_{3})]$). In this last case the universe
will end in a Big Rip singularity, where~$F,\dot{H}\rightarrow\infty$
and~$R\rightarrow0$. A Big Crunch ($F\rightarrow-\infty$) is not
possible since it contradicts~$R\rightarrow\infty$.

\subsection{\label{sub:B2}$\boldsymbol{\Lambda=\Lambda_{0}(L_{0}+L_{1}\frac{\mu^{2}}{\mu_{0}^{2}})}$,
$\boldsymbol{\mu=R^{-1}}$}

\begin{figure}
\begin{centering}\subfigure[]{\includegraphics[clip,width=0.48\columnwidth,keepaspectratio]{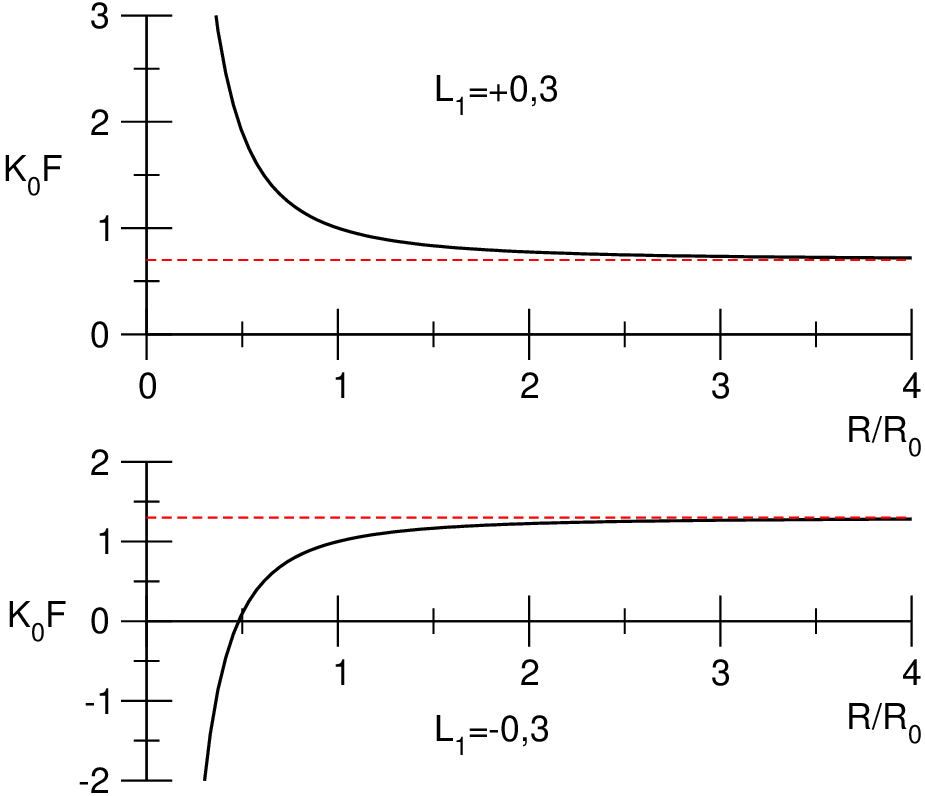}}\hfill{}\subfigure[]{\includegraphics[clip,width=0.48\columnwidth,keepaspectratio]{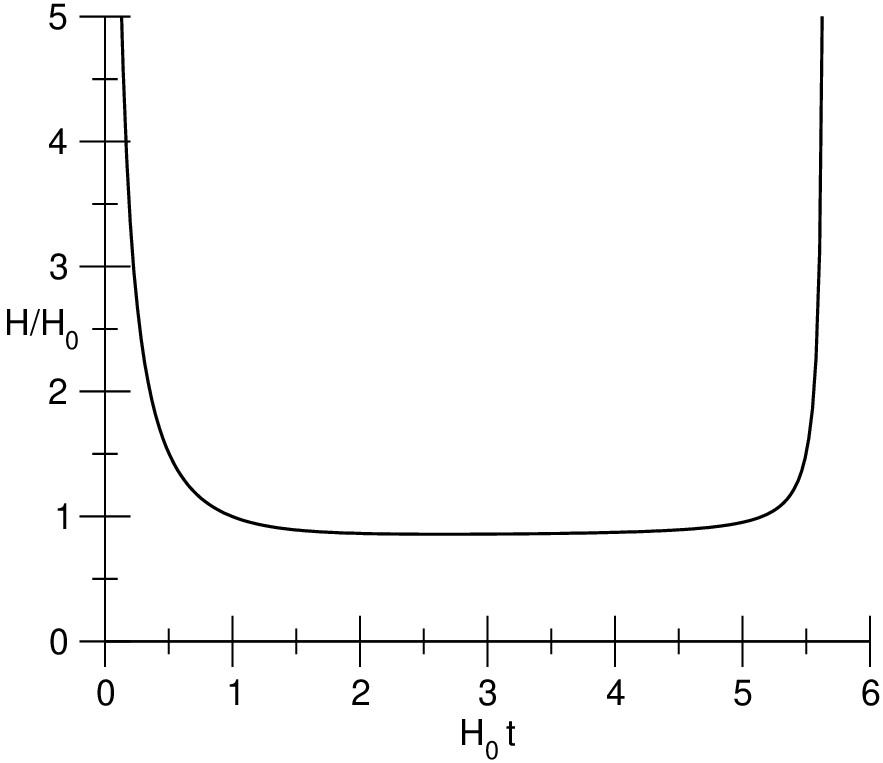}}\vspace*{-0.2cm}\par\end{centering}

\caption{\label{cap:B2-KoF-BR}Corresponding to Sec.~\ref{sub:B2} we plot
(a) the function $K_{0}F=L_{0}-+L\frac{R_{0}^{2}}{R^{2}}$ for $L_{1}=\pm0,3$
and (b) the Hubble scale $H(t)$ for~$L_{1}=+0,1$;~$R_{0}=1,1$;~$\Omega_{\Lambda0}=0,73$
and~$Q=\frac{1}{2}$ (dust), where a Big Rip singularity occurs.}
\end{figure}

In the de~Sitter limit,~$\dot{H}=0$ and~$R\rightarrow R_{\textrm{e}}=H_{\textrm{e}}^{-1}$,
we find the solution \[
R_{\textrm{e}}^{2}=(H_{0}^{-2}\Omega_{\Lambda0}^{-1}-L_{1}R_{0}^{2})/L_{0},\]
 but it is unstable for all possible values of~$L_{1}$, which follows
from \[
\left.\frac{\textrm{d}(K_{0}\dot{F})}{\textrm{d}(K_{0}F)}\right|_{\textrm{dS}}=R_{\textrm{e}}H_{0}^{2}L_{0}\Omega_{\Lambda0}>0.\]
 For~$L_{1}<0$ our numerical calculations did not yield any solution
that is compatible with Eqs.~(\ref{eq:Scale2}) and~(\ref{eq:MainEqu}).
This is also true in a certain parameter range for~$L_{1}>0$, where
otherwise a Big Rip singularity occurs, where $F,\dot{H}\rightarrow\infty$
and~$R\rightarrow0$, see Fig.~\ref{cap:B2-KoF-BR}. A Big Crunch
is not possible since~$K_{0}F$ is bounded from below.%

\subsection{\label{sub:C2}$\boldsymbol{\Lambda G=\Lambda_{0}G_{0}\frac{\mu^{2}}{\mu_{0}^{2}}}$,
$\boldsymbol{\mu=R^{-1}}$}

Here, one cannot find a prediction for~$R_{\textrm{e}}$ in the de~Sitter
limit ($\dot{H}=0$, $R\rightarrow R_{\textrm{e}}=H_{\textrm{e}}^{-1}$)
since Eq.~(\ref{eq:MainEqu}) only leads to the constraint~$R_{0}^{2}=1/(H_{0}^{2}\Omega_{\Lambda0})$.
To find solutions corresponding to other values of~$R_{0}$ we impose
the ansatz~$a(t)=a_{0}(t-t_{1})^{n}$ with~$a_{0},n,t_{1}=\textrm{const}$.
The Hubble scale is given by~$H=n(t-t_{1})^{-1}$ which implies due
to~$H_{0}>0$ that~$t_{1}<t$ for~$n>0$ and~$t_{1}>t$ for~$n<0$,
respectively. For~$0<n<1$ the event horizon does not exist, in the
other cases~$R$ can be calculated exactly via Eq.~(\ref{eq:Scale2}):\[
R(t)=a(t)\int_{t}^{t_{\textrm{E}}}\frac{\textrm{d}t^{\prime}}{a(t^{\prime})}=(t-t_{1})^{n}\cdot\frac{(t_{\textrm{E}}-t_{1})^{(1-n)}-(t-t_{1})^{(1-n)}}{1-n}.\]
 Note that for negative~$n$ there is a Big Rip singularity at~$t\rightarrow t_{1}$,
where~$a(t)$ diverges and the universe ends: $t_{\textrm{E}}=t_{1}$.
For~$n>1$ the scale factor~$a(t)$ describes power-law acceleration
and there is no future singularity which means~$t_{\textrm{E}}\rightarrow\infty$.
As a result, both cases yield \[
R(t)=\frac{(t-t_{1})}{n-1}.\]
Thus Eq.~(\ref{eq:MainEqu}) can be solved exactly, which determines
the constant~$n$, see Fig.~\ref{cap:C2-C3-n12Vonx}:%
\begin{figure}
\begin{centering}\includegraphics[clip,width=0.4\columnwidth,keepaspectratio]{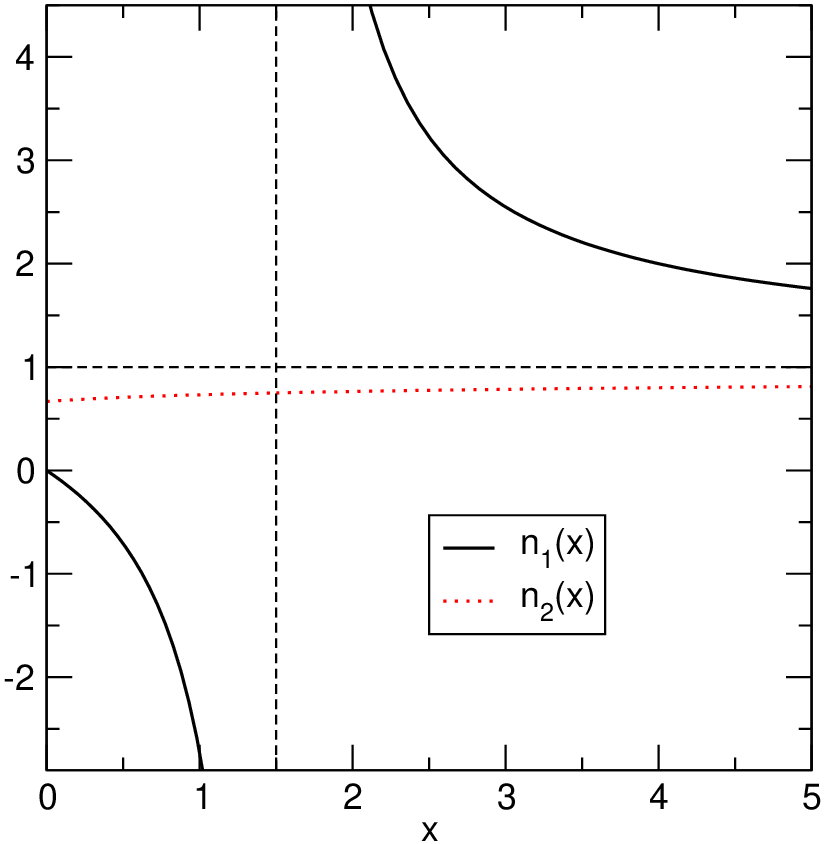}\par\end{centering}

\caption{\label{cap:C2-C3-n12Vonx}The exponents~$n_{1,2}(x)$ from Eq.~(\ref{eq:C2-C3-n12Vonx}),
which occur in the power-law expansion scale factor~$a(t)=a_{0}(t-t_{1})^{n}$,
are shown. In Sec.~\ref{sub:C2} we find $n=n_{1}(R_{0}^{2}/K_{0})$,
whereas $n=n_{2}(T_{0}^{2}/K_{0})$ in Sec.~\ref{sub:C3}.}
\end{figure}
\begin{eqnarray}
n_{1,2}(x) & = & \frac{(x-\frac{1}{2})\pm\sqrt{(x-\frac{1}{2})^{2}-x(x-(Q+1))}}{x-(Q+1)}\label{eq:C2-C3-n12Vonx}\\
\textrm{with}\,\, x & := & \frac{R_{0}^{2}}{K_{0}}=R_{0}^{2}H_{0}^{2}\Omega_{\Lambda0}(Q+1).\end{eqnarray}
For~$Q>0$ we find~$(Q+1)^{-1}<n_{2}<1$ for all positive values
of~$x$, therefore~$n_{2}$ can be dropped as the event horizon
does not exist. In the case~$R_{0}^{2}<1/(H_{0}^{2}\Omega_{\Lambda0})$
the constant~$n=n_{1}$ is negative leading to a Big Rip singularity
at the time~$t=t_{1}$, and~$R_{0}^{2}>1/(H_{0}^{2}\Omega_{\Lambda0})$
implies a positive~$n$ corresponding to power-law acceleration.

\subsection{\label{sub:A3}$\boldsymbol{\Lambda=\Lambda_{0}(1-q_{1}\ln\frac{\mu}{\mu_{0}})}$,
$\boldsymbol{\mu=T^{-1}}$}

Solving Eq.~(\ref{eq:MainEqu}) for this choice of~$\Lambda$ and~$\mu$
can be done approximately. First we propose the ansatz~$a(t)=a_{0}\exp(u^{2}(t-t_{1})^{2})$
for the scale factor with the constants~$a_{0}$,~$u$ and~$t_{1}$.
Therefore\[
H=2u^{2}(t-t_{1}),\,\,\,\,\frac{\ddot{a}}{a}=4u^{4}(t-t_{1})^{2}+2u^{2},\,\,\,\, F=2u^{2}+4u^{4}(Q+1)(t-t_{1})^{2},\]
and the particle horizon radius at late times reads\begin{eqnarray*}
T & = & \exp(u^{2}(t-t_{1})^{2})\int_{t_{x}}^{t}\exp(-u^{2}(t^{\prime}-t_{1})^{2})\textrm{d}t^{\prime},\,\,\,\, t>t_{x}=\textrm{const.}\\
 & = & \exp(u^{2}(t-t_{1})^{2})\cdot\left[\frac{\sqrt{\pi}}{2u}\textrm{erf}(u(t-t_{1}))\right]_{t_{x}}^{t}.\end{eqnarray*}
For~$t\rightarrow\infty$ the series expansion of the square brackets
reads~$C+\mathcal{O}(t^{-1})$ with~$C>0$ and thus~$T\approx C\exp(u^{2}(t-t_{1})^{2})$.
In this limit the given ansatz for~$a(t)$ solves~$K_{0}F=1-q_{1}\ln\frac{T_{0}}{T}$
exactly and determines the constant~$u^{2}=\frac{1}{4}q_{1}H_{0}^{2}\Omega_{\Lambda0}$.
Therefore we have found an approximate late-time solution in the case~$q_{1}>0$.

\begin{figure}
\begin{centering}\includegraphics[clip,width=0.48\columnwidth,keepaspectratio]{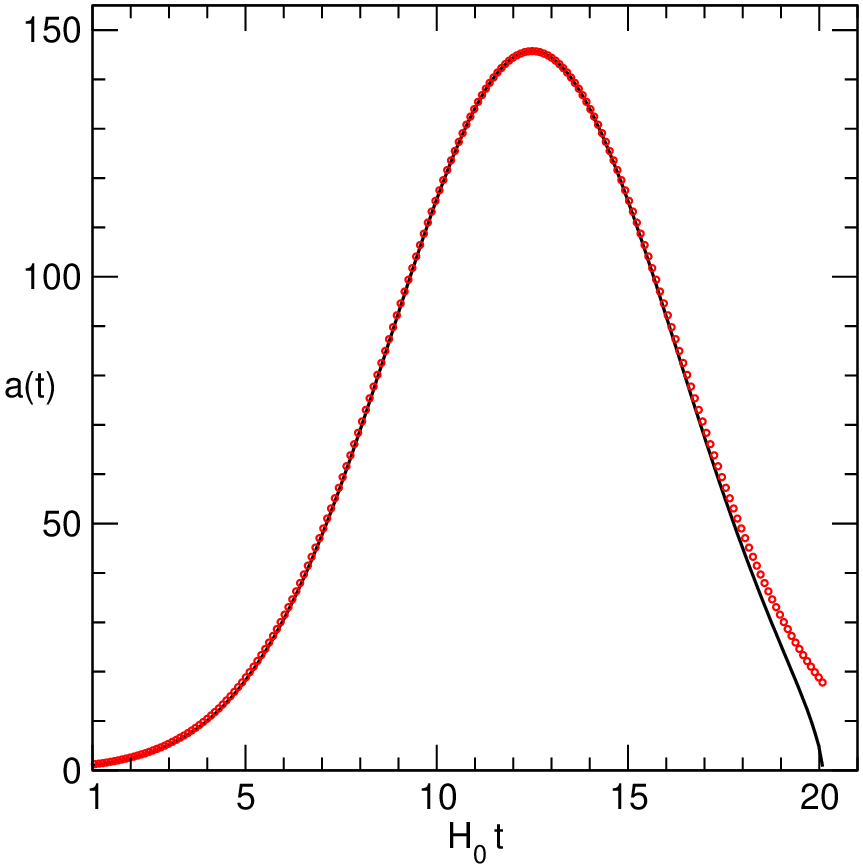}\hfill{}\includegraphics[clip,width=0.48\columnwidth,keepaspectratio]{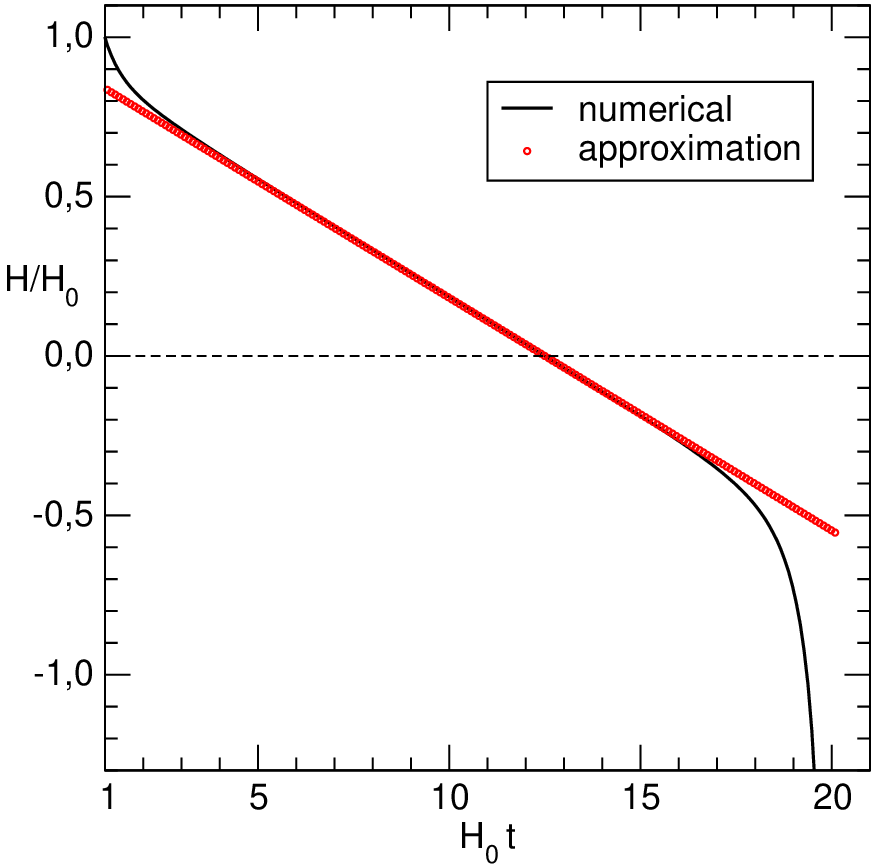}\par\end{centering}

\caption{\label{cap:A3-q1neg-a-H}Here, the numerical and the approximate
solutions for the scale factor~$a(t)$ and the Hubble scale~$H(t)$
from Sec.~\ref{sub:A3} are shown for negative~$q_{1}=-0,2$. Furthermore,
we used~$\Omega_{\Lambda0}=0,73$; $T_{0}H_{0}=3$ and~$Q=\frac{1}{2}$
(dust). In this case a Big Crunch singularity will occur in the future.
Note that the analytical approximation works very well except near
the future singularity.}
\end{figure}

The numerical solutions in the case~$q_{1}<0$ exhibits a future
Big Crunch singularity. At times sufficiently before this event the
scale factor can be well approximated by the ansatz~$a(t)=a_{0}\exp(-u^{2}(t-t_{1})^{2})$,
which implies~$F=-2u^{2}+4u^{4}(Q+1)(t-t_{1})^{2}$ and\[
T=\exp(-u^{2}(t-t_{1})^{2})\cdot\left[\frac{\sqrt{\pi}}{2iu}\textrm{erf}(iu(t-t_{1}))\right]_{t_{x}}^{t},\,\,\,\, t>t_{x}=\textrm{const}.\]
We expand the square bracket term around~$t\rightarrow t_{1}$, where
the scale factor is maximal, and find~$T\approx C\exp(-u^{2}(t-t_{1})^{2})$
with~$C>0$. Solving~$K_{0}F=1-q_{1}\ln\frac{T_{0}}{T}$ leads to~$u^{2}=-\frac{1}{4}q_{1}H_{0}^{2}\Omega_{\Lambda0}$
in accordance with~$q_{1}<0$. The numerical and approximate solutions
are shown in Fig.~\ref{cap:A3-q1neg-a-H}.%

\subsection{\label{sub:B3}$\boldsymbol{\Lambda=\Lambda_{0}(L_{0}+L_{1}\frac{\mu^{2}}{\mu_{0}^{2}})}$,
$\boldsymbol{\mu=T^{-1}}$}

In the late-time de~Sitter limit,~$a(t)\propto\exp(H_{\textrm{e}}t)$,
the particle horizon radius has the asymptotic form\[
T\propto\exp(H_{\textrm{e}}t)\rightarrow\infty\,\,\,\,\textrm{for}\,\,\,\, t\rightarrow\infty,\]
which follows from Eq.~(\ref{eq:T-dS}). This means that~$K_{0}F\rightarrow L_{0}$
and thus determines~$H_{\textrm{e}}^{2}=H_{0}^{2}\Omega_{\Lambda0}L_{0}$.
To verify the stability of this solution we calculate\[
K_{0}\dot{F}=-2L_{1}\frac{\dot{T}}{T}\cdot\frac{T_{0}^{2}}{T^{2}}=-2L_{1}\frac{T_{0}^{2}}{T^{2}}\left(H+\frac{1}{T}\right)\!\!.\]
At the initial time (today)~$K_{0}\dot{F}$ is negative for~$L_{1}>0$
and $K_{0}F$ keeps on approaching the de~Sitter limit unless a sign
change in~$K_{0}\dot{F}$ would occur. But this requires a negative~$H$,
which is only possible if~$K_{0}F=\dot{H}+(Q+1)H^{2}<0$ at some
time. Since~$K_{0}F\ge L_{0}>0$ we conclude that the de~Sitter
limit will be reached always. For~$L_{1}<0$ the initial values of~$K_{0}F$
and~$K_{0}\dot{F}$ are positive and~$K_{0}F$ approaches the de~Sitter
limit. Again, a sign change in~$K_{0}\dot{F}$ is only possible for~$H<0$,
which requires~$K_{0}F<0$ at some time. These conditions cannot
be realised because~$K_{0}F>0$ as long as~$K_{0}\dot{F}>0$. In
summary all reasonable values of~$L_{1}$ lead to a stable de~Sitter
final state.

\subsection{\label{sub:C3}$\boldsymbol{\Lambda G=\Lambda_{0}G_{0}\frac{\mu^{2}}{\mu_{0}^{2}}}$,
$\boldsymbol{\mu=T^{-1}}$}

By using at late times the ansatz~$a(t)=a_{0}(t-t_{1})^{n}$ with~$a_{0},t,n=\textrm{const.}$,
we find\[
F=(n^{2}(Q+1)-n)(t-t_{1})^{-2},\]
 and the particle horizon radius is given by\begin{equation}
T=a_{0}(t-t_{1})^{n}\left[\int_{0}^{t_{0}}\frac{\textrm{d}t^{\prime}}{a(t^{\prime})}+\int_{t_{0}}^{t}\frac{\textrm{d}t^{\prime}}{a(t^{\prime})}\right]=\frac{t-t_{1}}{1-n}+(t-t_{1})^{n}\left[T_{0}-\frac{(t_{0}-t_{1})^{1-n}}{1-n}\right]\!\!,\label{eq:C3-T}\end{equation}
where~$T_{0}=T(t_{0})>0$ depends on the past evolution of the scale
factor. In the case~$n<0$ a positive Hubble scale~$H=n(t-t_{1})^{-1}$
requires~$t,t_{0}<t_{1}$, which implies~$T<0$ and thus the non-existence
of the particle horizon. For~$0<n<1$ the second term in Eq.~(\ref{eq:C3-T})
can be neglected at late times, and Eq.~(\ref{eq:MainEqu}) can be
solved exactly, leading to \begin{eqnarray*}
n_{1,2}(x) & = & \frac{(x-\frac{1}{2})\pm\sqrt{(x-\frac{1}{2})^{2}-x(x-(Q+1))}}{x-(Q+1)}\\
\textrm{with}\,\, x & := & \frac{T_{0}^{2}}{K_{0}}=T_{0}^{2}H_{0}^{2}\Omega_{\Lambda0}(Q+1).\end{eqnarray*}
This is the same equation as in Sec.~\ref{sub:C2} with~$R_{0}$
replaced by~$T_{0}$, see also Fig.~\ref{cap:C2-C3-n12Vonx}. Here,
solution~$n_{2}$ is the physical one because of~$(Q+1)^{-1}<n_{2}<1$,
that corresponds to a decelerating universe for~$t\rightarrow\infty$.
In comparison with Sec.~\ref{sub:C2} the identification of the particle
horizon radius as renormalisation scale yields a complementary cosmological
behaviour.

\section{\label{sec:Conclusions}Conclusions}

In this work we have studied the effect of a variable CC on the late-time
evolution of the universe. Thereby the variability of the CC originates
from RGEs that are provided by certain quantum theories, leading to
a scale-dependence of the CC. By identifying the renormalisation scale
with cosmologically relevant scales, the CC becomes effectively time-dependent.
In contrast to many other models of DE, here, the EOS is still that
of a classical CC, which implies a non-trivial scaling behaviour of
the cosmological matter content. Concretely, we have investigated
three RGEs for the CC and three choices for the renormalisation scale:
the Hubble scale and the inverse radii of the cosmological event horizon
and particle horizon. For one RGE we have also included the renormalisation
group running of Newton's constant. In all cases we have determined
the possible final states of the universe and discussed their properties.
Depending on the choice of the initial conditions and parameters,
we have found at late times asymptotic de~Sitter solutions, accelerating
and decelerating power-law solutions and super-exponential expansion
laws of the type~$a(t)\sim\exp(ct^{2})$. Additionally, we have encountered
future singularities of the Big Rip and Big Crunch type. However,
some solutions should be treated with caution, since they imply physically
questionable results like negative energy densities. Also, the cosmological
behaviour near future singularities might get altered by additional
effects, that have not been taken into account in this work. It was
shown, for instance, in Refs.~\cite{BRip-QuCorr} that quantum effects
are able to prevent or weaken future singularities. Moreover, higher-derivative
terms in the gravitational action and a time-dependent Newton's constant
could potentially change the solutions. 

Nevertheless, we stay in this analysis on the basic level and use
the results we found as an indicator for the (in)stability of the
cosmological fate. Indeed, we have found regular solutions in many
cases, which can be considered as realisable in nature. In this sense,
we have found de~Sitter solutions in Secs.~\ref{sub:A1}, \ref{sub:B1},
\ref{sub:A2} and~\ref{sub:B3}. Also the power-law solutions can
be called regular, they appear for all cases with the scaling law~(\ref{eq:RGE3})
in Secs.~\ref{sub:C1}, \ref{sub:C1} and~\ref{sub:C3}. Moreover,
we have found super-exponential expansion laws in Sec.~\ref{sub:A3}.
Big Rip type future singularities exist for all cases with the event
horizon as renormalisation scale, see Secs.~\ref{sub:A2}, \ref{sub:B2}
and~\ref{sub:C2}. Big Crunch solutions, on the other hand, occur
only for the scaling law~(\ref{eq:RGE1}) as described in Secs.~\ref{sub:A1},
\ref{sub:A2} and \ref{sub:A3}. Finally, a comprehensive overview
of our results is given in Tab.~\ref{tab:Final-States}. Since some
solution types occur only for certain combinations of scaling laws
and scales, our analysis helps to discriminate between the different
cases and discover the nature of DE.%
\begin{table}
\begin{centering}\begin{tabular}{|l||c|c|c|}
\hline 
&
$\Lambda=\Lambda_{0}(1-q_{1}\ln\frac{\mu}{\mu_{0}})$&
$\Lambda=\Lambda_{0}(L_{0}+L_{1}\frac{\mu^{2}}{\mu_{0}^{2}})$&
$\Lambda G=\Lambda_{0}G_{0}\frac{\mu^{2}}{\mu_{0}^{2}}$\tabularnewline
\hline
\hline 
$\mu=H$&
\texttt{dS}, \texttt{BC}&
\texttt{dS}&
\texttt{P}\tabularnewline
\hline 
$\mu=R^{-1}$&
\texttt{dS}, \texttt{BR}, \texttt{BC}&
\texttt{BR}&
\texttt{P}, \texttt{BR}\tabularnewline
\hline 
$\mu=T^{-1}$&
$\exp(\pm t^{2})$, \texttt{BC}&
\texttt{dS}&
\texttt{P}\tabularnewline
\hline
\end{tabular}\par\end{centering}

\caption{\label{tab:Final-States}This table shows all types of cosmological
final states that we have found in Sec.~\ref{sec:Late-time} for
the all combinations of scaling laws for the CC~$\Lambda$ (and~$G$
for the third law) from Sec.~\ref{sec:Scaling-laws} and the choices
for the renormalisation scale~$\mu$ of Sec.~\ref{sec:Choice-scale}.
Here, \texttt{dS} denotes de~Sitter solutions, \texttt{P} power-law
solutions, \texttt{BR} Big Rip and \texttt{BC} Big Crunch future singularities,
respectively. With~$\exp(\pm t^{2})$ we mean the late-time behaviour
of the scale factor, where~$a\propto\exp(-t^{2})$ finally leads
to a Big Crunch.}
\end{table}
In a future work, we would like to discuss the consequences of a running
CC for the early universe, too.

\section*{Acknowledgements}

I wish to thank H.~\v{S}tefan\v{c}i\'{c} for useful discussions.
This work was supported by the {}``Sonderforschungsbereich 375 für
Astroteilchenphysik der Deutschen Forschungsgemeinschaft''.

\end{document}